\documentclass[10pt]{article}
\usepackage{amsmath}
\usepackage{graphicx}
\usepackage{amsfonts}
\usepackage{amssymb}
\usepackage{epsf}
\usepackage{latexsym}

\def\be{\begin{equation}}
\def\ee{\end{equation}}
\def\bea{\begin{eqnarray}}
\def\eea{\end{eqnarray}}

\def\beq{\begin{equation}}
\def\eeq{\end{equation}}
\def\bea{\begin{eqnarray}}
\def\eea{\end{eqnarray}}
\def\pa{\partial}

\def\5Star{\mbox{\Large$\star$}}

\def\sumi3{\sum\mbox{}_{\mbox{}_{\mbox{\scriptsize $i$=1}}}^3}

\def\sumj3{\sum\mbox{}_{\mbox{}_{\mbox{\scriptsize $j$=1}}}^3}
\def\sumk3{\sum\mbox{}_{\mbox{}_{\mbox{\scriptsize $k$=1}}}^3}

\def\mk{\mbox{k}}

\def\mA{\mbox{A}}

\def\se{\mbox{\scriptsize e}}

\def\so{\mbox{\scriptsize o}}

\def\st{\mbox{\scriptsize t}}

\def\sB{\mbox{\scriptsize B}}

\def\sH{\mbox{\scriptsize H}}

\def\sT{\mbox{\scriptsize T}}

\def\eph{\mbox{\scriptsize eph}}
 
\def\eph(B){\mbox{\scriptsize em(JBB)}}

\def\eph(B){\mbox{\scriptsize emergent(JBB)}}

\begin{document}
\begin{center}

\large{\bf KERR--NEWMAN BLACK HOLE THERMODYNAMICAL STATE SPACE:} \normalsize

\vspace{0.1in}

\large{\bf BLOCKWISE COORDINATES}\normalsize 

\vspace{0.1in}

\normalsize

\vspace{0.1in}

{\large \bf Edward Anderson$^*$}

\vspace{.2in}

{\large {\em DAMTP Cambridge \normalsize}}.

\vspace{.2in}

\end{center}

\begin{abstract}

A coordinate system that blockwise-simplifies the Kerr--Newman black hole's thermodynamical state space Ruppeiner metric geometry is constructed, 
with discussion of the limiting cases corresponding to simpler black holes.
It is deduced that one of the three conformal Killing vectors of the Reissner--Nordstr\"{o}m and Kerr cases (whose thermodynamical state space metrics 
are 2 by 2 and conformally flat) survives generalization to the Kerr--Newman case's 3 by 3 thermodynamical state space metric. 

\end{abstract}

\noindent


\vspace{0.1in}
 
\noindent $^*$ ea212@cam.ac.uk

\section{Introduction}

This brief paper concerns the Riemannian geometry of thermodynamical state spaces, which is of wider interest as one possible means of studying phase transitions.
I follow Ruppeiner's approach \cite{Ruppeiner79, Ruppeiner95, Ruppeiner08}; see e.g. \cite{Weinhold, FGK, ABP, Shen, OtherStyle, Quevedo} and Sec 4 for other approaches. 
I consider the black holes case of this, for which the thermodynamical fluctuation formalism itself has difficulties concerning stability about a maximum in the total entropy, 
so that an alternative to it is desirable.

Ruppeiner \cite{Ruppeiner79, Ruppeiner08} considers the fluctuation expansion
\beq
\Delta S_{\sT\so\st} = F_{\mu} \Delta X^{\mu} + F_{e\mu}\Delta X^{\mu}_e + \frac{1}{2}\frac{\pa F_{\mu}}{\pa X^{\nu}_e}\Delta X^{\mu}\Delta X^{\nu} 
+ \frac{1}{2}\frac{\pa F_{e\mu}}{\pa X^{\nu}_e}\Delta X^{\mu}_e \Delta X^{\nu}_e + ... \mbox{ } .  
\eeq
Here, $S_{\sT\so\st} = S_{\sB\sH} + S_{\se}$ is the total entropy of the universe,  
the standard black hole thermodynamical variables are denoted by $X^{\mu}$ 
[e.g. ($X^1$, $X^2$, $X^3$) := ($M$, $J$, $Q$) for the Kerr--Newman black hole], $S$ is entropy, and 
\beq
F_{\mu} = \pa S/\pa X^{\mu} \mbox{ } .
\eeq
The subscript `e' stands for the `environment' `external' to the black hole.  
For a very large, extensive environment, 
\beq
\Delta S_{\sT\so\st} = - \frac{1}{2}g_{\mu\nu}\Delta X^{\mu} \Delta X^{\nu}  
\eeq
transforms as a scalar due to depending only on the initial and final thermodynamical states.  
One can view the corresponding Ruppeiner geometry as an {\it information geometry of Fisher type} \cite{Fisher}. 
The associated Ricci scalar curvature is then argued by Ruppeiner to be a measure of interaction strength.  
Ruppeiner computed the form of this for the Kerr--Newman black hole solution's thermodynamical state space in e.g. \cite{Ruppeiner08}.

Further coverage of the Kerr--Newman black hole case is the subject of this paper. 
This is a case of particular importance due to this 4-$d$ black hole \cite{KNitself} likely being astrophysically realized (at least to very good approximation as a model and 
for very small values of the charge parameter). 
This is due to its relevant and not excessively high symmetry,  uniqueness theorems leading to it, and its capacity to arise as the end-product of the evolution of somewhat 
more irregular configurations.
See e.g. \cite{4dBH, Shen, ABP, Quevedo} for previous work on the geometry of thermodynamic state spaces for 4-$d$ black holes.  
In the present paper, a mathematically superior coordinate system for the Ruppeiner thermodynamical state space geometry for the Kerr--Newman black hole is found (Sec 3).
This is obtained via the following sequence of coordinate transformations. 
I take Davies coordinates \cite{4}, apply mass-homogenization and then apply a new coordinate transformation; 
I also present the overall composite transformation and consider limiting cases (Reissner--Nordstr\"{o}m, Kerr, extremal).
The coordinate system found is superior in particular by virtue of block-minimality.
Consequences of this include (Sec 4) 1) obtaining geometrical insights (showing using my coordinate system that 
1 of the 3 conformal Killing vectors of the limiting cases' thermodynamical state spaces survives the generalization to the Kerr--Newman case).  
2) Saving considerably in the computation of geometrical quantities by working with a metric that has the minimum number of nonzero components.

\section{Ruppeiner thermodynamical geometry in the Kerr--Newman case}

\noindent For the physically-valuable 4-$d$ Kerr--Newman solution, the entropy is given by \cite{Smarr}
\beq
S = \pi \, \mk_{\sB}[2M^2(1 + \bar{E}) - Q^2] \label{In-Ent}
\eeq
for
\beq
\mbox{`nonextremality' } , \mbox{ } \bar{E} \mbox{ }  := \sqrt{1 - Q^2/M^2 - J^2/M^4} \mbox{ } \mbox{ } .
\eeq
With these standard thermodynamic state variables as coordinates, the thermodynamical Ruppeiner metric is as in Fig 1a).  
The idea is then to blockwise-simplify this by applying coordinate transformations.

\section{Its blockwise simplification}

I first use a transformation well-known to Ruppeiner \cite{Ruppeiner08} and which goes at least as far back as to Davies \cite{4}: 
\beq
q := Q/M \mbox{ } , \mbox{ } \mbox{ } j := J/M^2 \mbox{ } ; 
\eeq
these are a particular case of dimensionless variables.  
I note that these then simplify the blockwise form of the thermodynamical Ruppeiner metric as in Fig 1b).  
I next apply a mass-homogenizing/$M$-dependence-freeing transformation  
\beq
{\cal M} := \mbox{ln}\,M  \mbox{ } , 
\eeq
so as to have a fully homogenized version of Davies-type coordinates.
This casts the metric in the form of Fig 1c).  
%
{            \begin{figure}[ht]
   \centering
\includegraphics[width=0.95\textwidth]{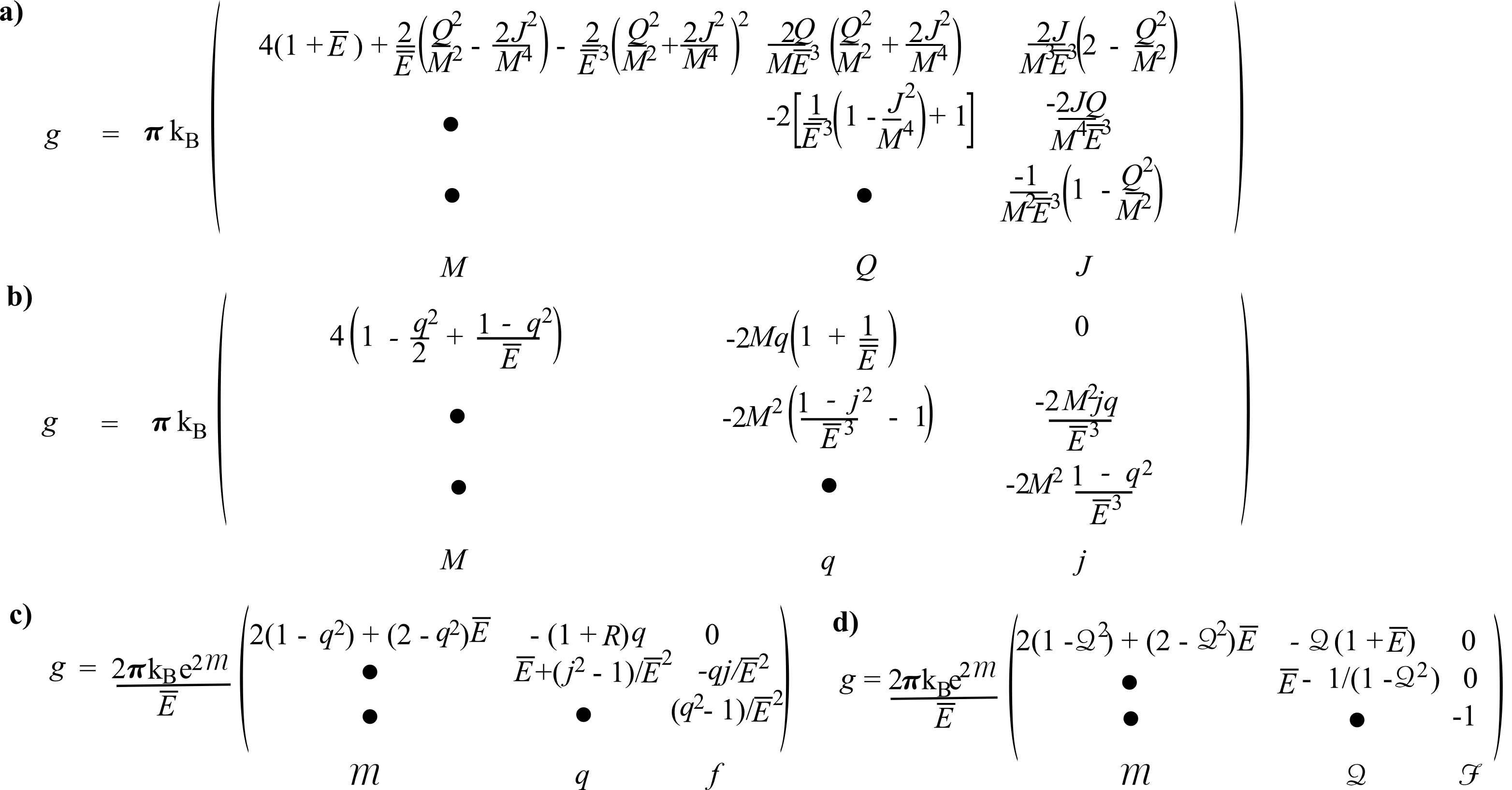}
\caption[Text der im Bilderverzeichnis auftaucht]{        \footnotesize{The Ruppeiner thermodynamic metric for the Kerr--Newman black hole 
in a) standard state variable coordinates, b) Davies coordinates, c) my mass-homogenized version of Davies coordinates and d) my block-minimal coordinates.}    }
\label{FigFirst} \end{figure}          }

Finally, I obtain a further cross-term removing coordinate transformation alongside a simplification for the isolated $1 \times 1$ block by solving 
a pair of differential equations, yielding the coordinate transformation
\beq
{\cal F} := \mbox{arcsin}\big( j/\sqrt{1 - q^2} \big)
\eeq
by which the metric is cast in the final blockwise-simplified form of Fig 1d).  
[I use ${\cal Q}:= q$ to have the same typeface for my final set of three blockwise-simplifying coordinates.]

Composing, the complete transformation is
\beq
{\cal M} = \mbox{ln}\,M                       \mbox{ } , \mbox{ } \mbox{ }
{\cal Q} = Q/M                                       \mbox{ } , \mbox{ } \mbox{ }
{\cal F} = \mbox{arcsin}\big( J/M\sqrt{M^2 - Q^2} \big) \mbox{ } , 
\eeq
with inverse 
\beq
M = \mbox{exp}{\cal M} \mbox{ } , \mbox{ } \mbox{ } 
Q = {\cal Q}\,\mbox{exp}\,{\cal M} \mbox{ } , \mbox{ } \mbox{ } 
J = \sqrt{1 - {\cal Q}^2}\, \mbox{exp}(2{\cal M})\,\mbox{sin}\,{\cal F} \mbox{ } .
\eeq
The form of $\bar{E}$ progresses as follows along the chain of coordinate transformations: 
\beq
\bar{E} = \sqrt{1 - Q^2/M^2 - J^2/M^4} = \sqrt{1 - q^2 - j^2} = \sqrt{1 - {\cal Q}^2}\,\mbox{cos}\,{\cal F} \mbox{ } .
\eeq

\section{Discussion and Applications}

\noindent {\bf 1) Limiting cases of physical interest} 
i) ${\cal F} = 0$ corresponds to $J = 0$, i.e. the Reissner--Nordstr\"{o}m black hole. 
For this case, the Ruppeiner geometry is flat, so it cannot encode this case's phase transition. 
This is a well-known limitation on the Ruppeiner thermodynamical geometry being an entire account of phase transitions. 
Some suggestions as regards how to include this case are i) to exclude the static electric energy \cite{Shen}. 
ii) This exclusion is an example of Legendre transformation, a type of transformation familiar from basic thermodynamics. 
Moreover, a case can be made for adopting thermodynamic state space metrics that are invariant under Legendre transformations \cite{Quevedo}.  
This illustrates that thermodynamical state space metrics remain an unfinished subject.  

\noindent ii) Holding $q$ constant gives a $2 \times 2$ diagonal block (here, Davies coordinates suffice to diagonalize).  
This includes Kerr for $q = 0$, c.f. \cite{Ruppeiner08}, and is also conformally flat.  

\noindent iii) The extremal values $\pm \pi/2$ of ${\cal F}$ (see Fig 2) correspond to $j^2 + q^2 = 1$, i.e. the extremal Kerr--Newman black hole. 
Here, this paper's coordinate system breaks down; moreover the thermodynamic state space in this case comes out straightforwardly in 
$x = 2M$, $t = \sqrt{2}Q$ coordinates as flat ($2 \times 2$ Minkowski metric), so this case is easily elsewise covered.
%
{            \begin{figure}[ht]
   \centering
\includegraphics[width=0.7\textwidth]{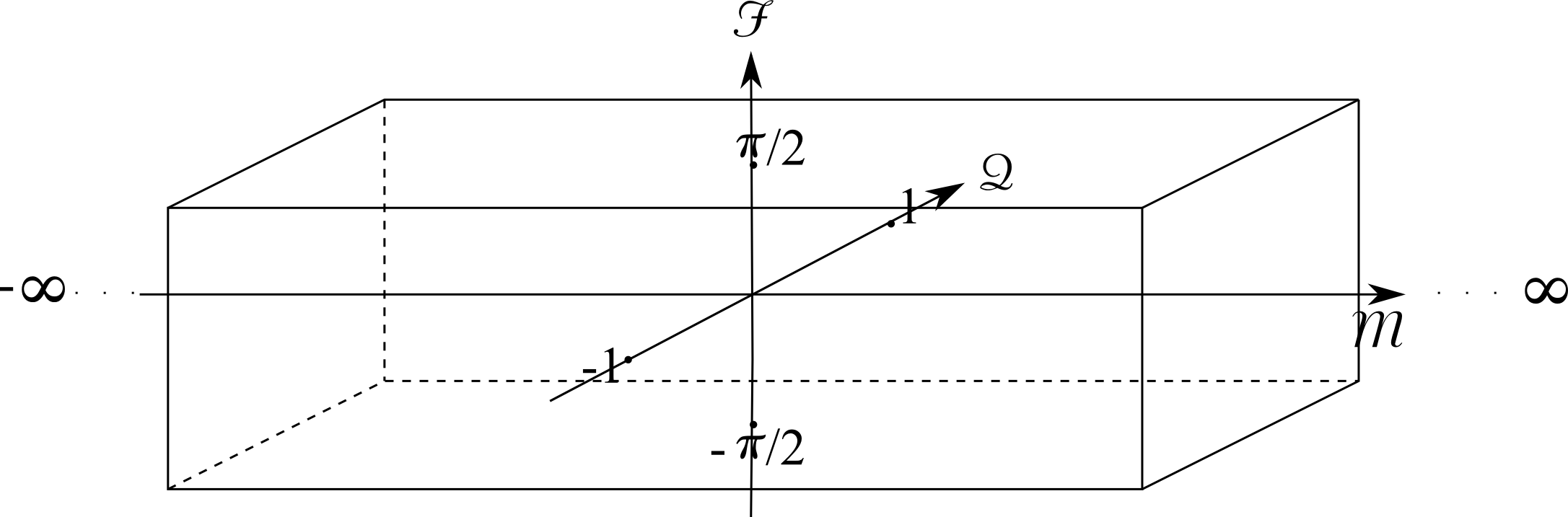}
\caption[Text der im Bilderverzeichnis auftaucht]{        \footnotesize{Shape of the Kerr--Newman thermodynamical state space in this paper's coordinates.  
The extremal values of {\cal F} are not covered by these coordinates.}    }
\label{FigLast} \end{figure}          }

\noindent {\bf 2) Note on the general value of block-minimality}.  
Block-minimality is a weakening of the concept of diagonality for cases in which diagonality itself cannot be attained; 
this paper's working does not extend to a removal of the last independent off-diagonal term.  
Well-known examples of non-diagonalizable metrics and useful block-minimal forms are as follows.  

\noindent 1) For $\mathbb{CP}^2$ geometry, Gibbons--Pope coordinates \cite{GiPo} are block-minimal and subsequently serve e.g. as useful cyclic coordinates for dynamics on 
$\mathbb{CP}^2$, in the analysis of conserved quantities on this space and in separating the Schr\"{o}dinger equation on this space \cite{QuadItoIII} and for a number of gravitational 
instanton applications.

\noindent 2) The most general Bianchi IX anisotropic models are non-diagonal; these are of value in modelling both dynamics (some of the more general cases of dynamical systems behaviour 
in \cite{Wainwright-and-Ellis}) and singularities (e.g. the BKL conjecture \cite{BKL}) in cosmology.  

\noindent 3) For Kerr and Kerr--Newman spacetimes, Boyer--Lindquist coordinates are block-minimal, and as such are widely useful in understanding rotational effects in black hole physics.  

\noindent 4) For the Legendre transformation invariant thermodynamical state space geometry of \cite{Quevedo} in the Kerr--Newman case also, the metric is also 
non-diagonal with a block-minimal form provided.
Moreover, this metric's block-minimal form is distinct from ours via its sole nonzero off-diagonal term being `$Q, J$' rather than `$M, Q$'. 
This distinction means that 4)'s block minimality aligns with the $M^2f(q, j)$ factorization of (\ref{In-Ent}) so that the Hessian of $f$ itself enters the expression for this metric.

\mbox{ } 

\noindent {\bf 3) Some applications of the present paper's case's block-minimality}.

\noindent A) Study of conformal Killing vectors (CKV's). Each of the Reissner--Nordstr\"{o}m and Kerr black hole thermodynamic state spaces have 3 CKV's by conformal flatness.
In the generalization to Kerr--Newman black hole thermodynamical state space, at least one CKV survives: $\pa/\pa{\cal M}$.  
The homogenizing transformation that unveiled this here does have a counterpart in the case of the simpler Kerr black hole thermodynamical state space itself. 
However, this kind of transformation was not arrived at in this case, probably because it is apparent in this case by other means that the 
Ruppeiner thermodynamical state space geometry is conformally flatness, by which one trivially obtains the CKV's in this case.     
This triviality also occurs for the Reissner--Nordstr\"{o}m black hole's Ruppeiner thermodynamic state space.  
The value of finding CKV's is additionally increased by these being shared by the Weinhold thermodynamical geometry \cite{Weinhold} since this is conformally related to the Ruppeiner one.  
%

\noindent B) Mathematical characterization of the geometry involved. 
It is not clear at first sight whether a geometry found in some physical application is already known in mathematics or in other areas of physics. 
This is because of coordinate dependence of line elements, and makes e.g. deciding whether one's endeavours have resulted in one finding a 4-metric that is a 
new solution to Einstein's equations is a nontrivial process \cite{MacCallum}.  
One has to consider a number of coordinate-invariant features of the geometry in order to determine this. 
The situation of geometrical characterization is similar in the case of 3-metrics such as the present paper's, which moreover arises in an interesting physical context 
as a thermodynamic state space for the most basic, very fundamental and most likely to be astrophysically realized family of black holes.  
Knowledge of the number and algebraic inter-relations between the (conformal) Killing vectors is useful in this direction, 
as is the computation of the full set of curvature invariants \cite{MacCallum}. 
The present paper's coordinate system not only readily sheds some information about the former 
(I do not recognise this geometry as occuring elsewhere in the mathematical or physical literature, but I have uncovered that it manifests at least a moderate amount of symmetry 
in the form of it possessing a conformal Killing vector).  
It also considerably simplifies the computation of the latter through starting with the least number of nonzero metric components.

\mbox{ }

\noindent {\bf Acknowledgements}.  
I thank Ingemar Bengtsson for pointing out the alignment in point 2.4 of the Discussion, albeit in the context that my blockwise minimal expression for the Ruppeiner thermodynamic 
state space metric for the Kerr--Newman black hole did not possess it.  
It was down to me to point out that, on the other hand the Legendre transformation invariant thermodynamic state space metric for the Kerr--Newman black hole {\sl does} 
possess this alignment, which could be helpful in the study of this other state space metric metric.  
I also thank Narit Pidokrajt, Malcolm MacCallum and the anonymous referees for various discussions and points.  
I acknowledge Grant FQXi-RFP3-1101 from the Foundational Questions Institute (FQXi) Fund, administered by 
Silicon Valley Community Foundation, Theiss Research and the CNRS, hosted with Marc Lachieze-Rey at APC.
Finally, this work was started whilst at Peterhouse and DAMTP, Cambridge, so I also acknowledge a Peterhouse Research Fellowship.


\end{document}